\pdfoutput=1

\documentclass[10pt,letterpaper]{article}
\usepackage[top=0.85in,left=2.75in,footskip=0.75in]{geometry}

\usepackage{changepage}

\usepackage[utf8]{inputenc}

\usepackage{textcomp,marvosym}

\usepackage{fixltx2e}

\usepackage{amsmath,amssymb}

\usepackage{cite}

\usepackage{nameref,hyperref}

\usepackage[right]{lineno}

\usepackage{microtype}
\DisableLigatures[f]{encoding = *, family = * }

\usepackage{rotating}

\usepackage{amsfonts}
\usepackage{amsmath}
\usepackage{enumerate}
\raggedright
\usepackage[aboveskip=1pt,labelfont=bf,labelsep=period,justification=raggedright,singlelinecheck=off]{caption}

\makeatletter
\renewcommand{\@biblabel}[1]{\quad#1.}
\makeatother

\date{}

\usepackage{lastpage,fancyhdr,graphicx}
\usepackage{epstopdf}
\fancyheadoffset[L]{2.25in}
\fancyfootoffset[L]{2.25in}

\begin{document}
\vspace*{0.35in}

\begin{flushleft}
{\Large
\textbf\newline{A geometric graph model of citation networks with linearly growing node-increment}
}
\newline
\\
Qi Liu\textsuperscript{1,*},
Zheng Xie\textsuperscript{1},
Enming Dong\textsuperscript{1},
Jianping Li\textsuperscript{1}
\\
\bigskip
\bf{1} College of Science, National University of Defense Technology, Changsha, Hunan, China
\bigskip

%
%





* liuqi@smail.nju.edu.cn

\end{flushleft}
\section*{Abstract}
Due to the fact that the numbers of annually published papers have witnessed a linear growth in some citation networks,
a geometric model is thus proposed to predict some statistical features of those networks, in which the academic influence scopes of the papers are denoted through specific  geometric areas related to time and space. In the model, nodes (papers) are uniformly and randomly  sprinkled onto a cluster of circles  of the Minkowski space whose centers are on the time axis.
Edges (citations) are linked according to an influence mechanism which indicates that an existing paper  will be cited by a new paper located in its influence zone. Considering the citations among papers in different disciplines, an interdisciplinary citation mechanism is added to the model in which some papers with a small probability of being chosen will cite some existing papers randomly and uniformly. Different from most existing models that only study the power-law tail of the in-degree distribution, this model also characterizes the overall in-degree distribution.  Moreover, it presents the description of some other important statistical characteristics of real networks, such as  in- and out-assortativity,  giant component and  clear community structure.  Therefore, it is reasonable to believe that a good example is provided in the paper to study real networks by geometric graphs.


\section*{Introduction}
The research of citation networks has drawn increasing attention and been applied to many fields\cite{ref1,ref2,ref3}. It can help scientists find useful academic papers\cite{liu1}, help inventors  find interesting patents\cite{liu2}, or help judges  discover relevant past judgements\cite{liu3}. The scientific citation networks considered in this paper are directed graphs, in which nodes represent papers, while edges stand for  the citation relationships between them. Since new papers can only cite the published papers\cite{liu12}, these graphs are acyclic.

Degree distribution is a fundamental research object of citation networks, and  a series of  models have been proposed to illustrate it. The Price model appears to be the first to discuss about cumulative advantage in the context of citation networks and their in-degree distributions\cite{liu4}.  The idea lies in that the rate at which a paper gets new citations should be proportional to the citations that it already has\cite{liu5,liu6}. This can lead to a power-law distribution according to the Price model\cite{liu7}. A copy mechanism in which a new node attaches to a randomly selected target node as well as all its ancestors has been proposed by Krapivsky et al\cite{liu8}. Based on their viewpoints, an author may be familiar with a few primary references and may simply copy the secondary references from the primary ones.
This rule also leads to a power-law distribution. In addition, the cumulative advantage is also known as the preferential attachment in other literatures\cite{liu9,liu10,liu11}. Jeong et al\cite{liu11} have measured  the rate at which nodes acquire links on four kinds of  real networks, and found that it depends on the node's degree. Their results offer direct quantitative support for the presence of preferential attachment.
Moreover, an investigation has been conducted by Eom et al\cite{liu12} on the microscopic mechanism for the evolution of citation networks by raising a linear preferential attachment with time-dependent initial attractiveness.
The model reproduces the tails of the in-degree distributions of citation networks and the phenomenon called ``burst": the citations received by papers increase rapidly in the early years since publication.
The above-mentioned models have studied the tail of the in-degree distribution only, while the two-mechanism model proposed by George et al\cite{liu7} characterizes the properties of the overall in-degree distributions.

With respect to the research of networks from the real world (e.g. citation networks), using random geometric graph has become a hot topic in recent years.
 Xie et al\cite{liu14}define the academic influence scope as a geometric area and present an influence mechanism, which means that an existing paper will be cited by a new paper located in its influence zone. Based on this mechanism, they  further propose the concentric circles model (CC model), which can fit the power-law tails of the in-degree distributions of the citation networks with the exponentially growing nodes.
 Nevertheless, the forepart of the in-degree distribution and the out-degree distribution can not be well fitted by this model.

\begin{figure}
\begin{adjustwidth}{-2.25in}{0in}
\centering
 \includegraphics[height=1.5in,width=6in,angle=0]{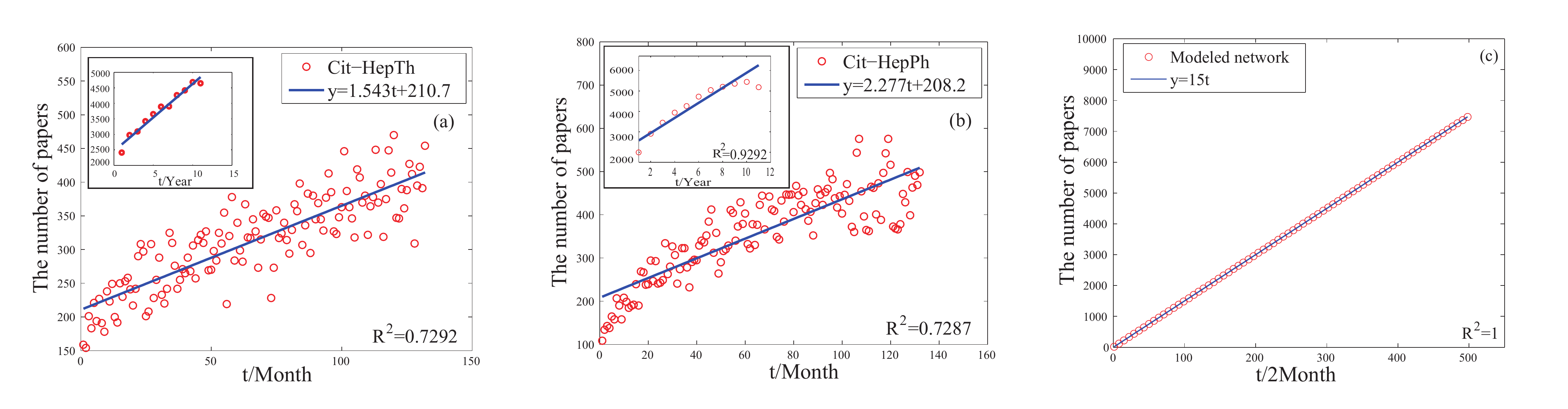}
\caption{\small\textbf{The changing trends of the monthly numbers of papers of the data in Table~\ref{table1}.} Panels (a,b) show the trends for the papers of Cit-HepTh and Cit-HepPh. Panel (c) shows the trend for the modeled network. They are fitted by linear functions. The coefficient of determination $(R^{2})$ is used to measure the goodness of fits.  }
\label{fig1}       
\end{adjustwidth}
\end{figure}

In reality, node-increment in many current citation networks enjoys a linear growth, e.g. Cit-HepPh,  Cit-HepTh\cite{liu15,liu16} (Fig~\ref{fig1}) and PNAS including articles published during 2000-2015 (shown in our later study). Therefore, a model with linearly growing node-increment is proposed. The edges in the model are still linked according to the influence mechanism,
whereas they are revised in that the influence scopes of papers are determined by their topics and ages (the time that has passed since publication).
Different from the previous models that only focus on the tails of in-degree distributions, the improved model can well predict the overall in-degree distributions of the empirical data well. In consideration of the citations among different disciplines in real citation networks, a mechanism that is referred to as the interdisciplinary citation mechanism is proposed. Under appropriate parameters, these mechanisms can reproduce a range of properties of citation networks, including the power-law tail of the out-degree distribution, giant component and clear community structure. Meanwhile, some other properties can also be obtained like the relationship between in-degree and local clustering coefficient as well as in- and out-assortativity. These results show that our model can be used as a medium to study the intrinsic mechanism of citation networks.
\begin{table}[!t]
\begin{adjustwidth}{-2.25in}{0in}
\renewcommand{\arraystretch}{1.3}
\setlength{\abovecaptionskip}{0pt}
\setlength{\belowcaptionskip}{10pt}
\caption{\small\textbf{The first two networks extract from arXiv which cover paper from January 1993 to April 2003 (124 months)  in high energy physics theory and in high energy physics phenomenology\cite{liu15,liu16}.} The last network is generated according to the generating process of the model, where parameters are $m=15$, $T=66$, $\beta_{0}=0.035$, $\lambda=0.001$, $S=66$, $p=1$, $\eta=2.5$, $\alpha=1.3$, $r=0.01$, $\xi=2.7$, $k_{0}=6$. In the header of the table, CC, AC, AC-In, AC-Out, PG and MO denote the clustering coefficient, the assortative coefficient, the in-assortative coefficient, the out-assortative coefficient, the node proportion of giant component and modularity, respectively.}
\centering

\begin{tabular}{  l  c  c  c  c  c c c c} \hline\hline
Networks&Nodes& Links&CC&AC& AC-In &AC-out &PG &MO\\ \hline
Cit-HepTh &27770 &352807 &0.165 &-0.030   &0.041 &0.096 &0.987 &0.650\\
Cit-HepPh &34546 &421578 & 0.149&-0.006   &0.077 &0.112 &0.999 &0.724\\
Modeled network &33165 &162080&  0.393 &-0.068 &0.316 &0.166 &0.970&0.967\\
\hline\hline
  \end{tabular}
  \label{table1}
  \end{adjustwidth}
\end{table}

The structure of this paper is as follows.  The model is described in Section 2. The degree distributions, clustering and assortativity are analyzed in Section 3 to Section 5, and finally  the conclusion is provided in the last section.

\section*{The model}
\begin{figure}
\begin{adjustwidth}{-2.25in}{0in}
\centering
 \includegraphics[height=2.3in,width=2.6in,angle=0]{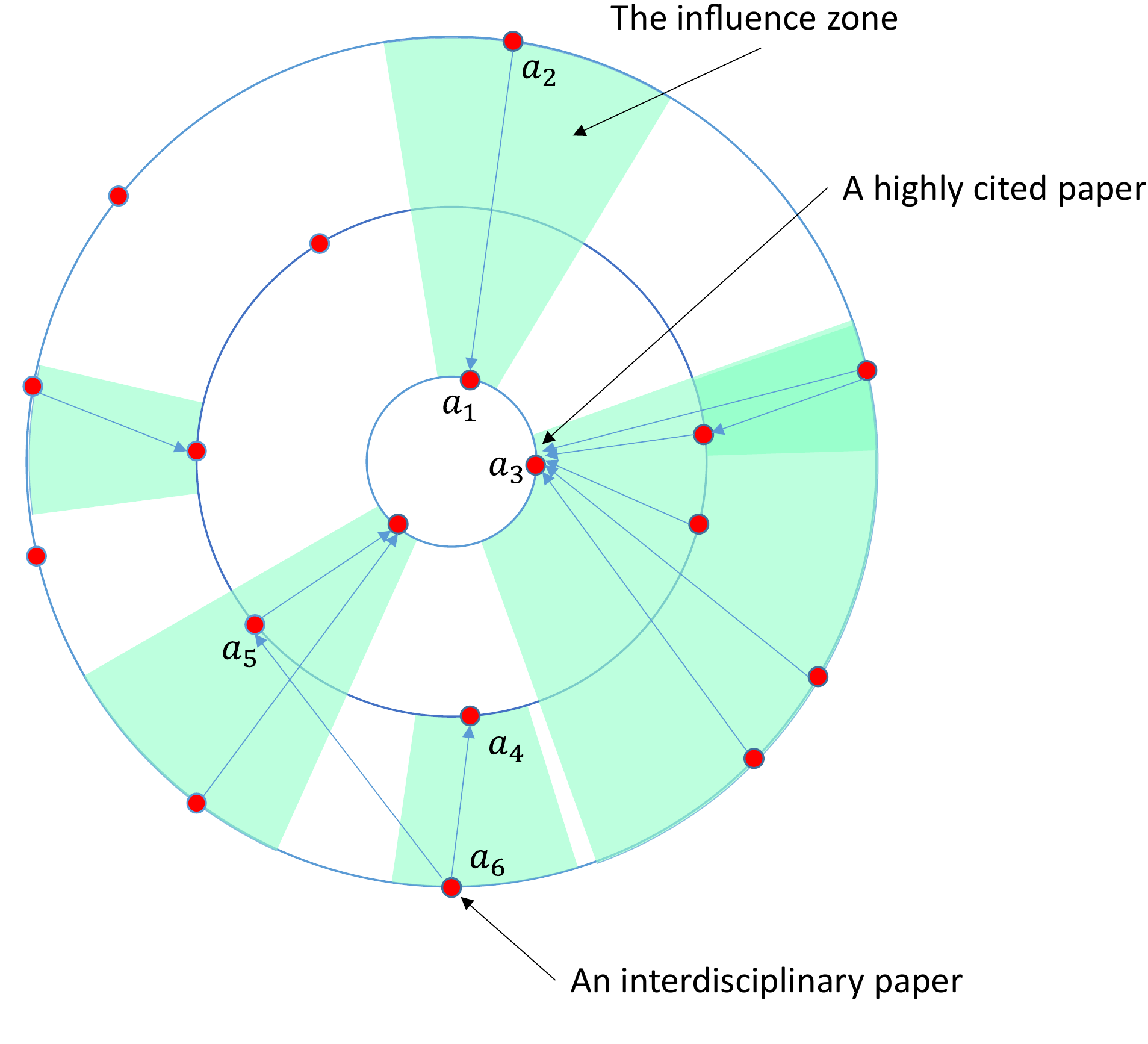}
\caption{\small \textbf{Diagram of the model}. The red dots on each circle denote the papers published in the same issue, the number of which grows linearly. The cyan areas represent the influence zones of nodes. Diversities between the academic influences are expressed by different zonal sizes, e.g. the influence zone of the highly cited paper $a_{3}$ is bigger than other's. The edges in the model are linked according to the influence mechanism and the interdisciplinary citation mechanism: node $a_{2}$ belongs to the zone of node $a_{1}$, then $a_{2}$ cites $a_{1}$; node $a_{6}$ is an ``interdisciplinary paper", so it could cite node $a_{5}$,  even though $a_{6}$ does not belong to the influence zone of $a_{5}$.    }
\label{dia}
\end{adjustwidth}
\end{figure}
Since many journals and databases publish papers monthly or yearly and papers in the same issue cannot cite each other normally, models like the Price model or the copy model that publish one paper at each time step do not consider the growing trends of papers. Xie et al\cite{liu14} pay attention to the citation networks  in which the annual numbers of papers grows exponentially, such as  the citation network collected by Tang et al\cite{liu17} for papers (which are published in the period from 1936-01-01 to 2013-09-29) in DBLP dataset.  However, in some real citation networks (e.g. Cit-HepPh and Cit-HepTh), the monthly or annual numbers of papers published   grow linearly (Fig~\ref{fig1}).  For purpose of  the evolution and features of these networks, a geometric graph model,  in which the  node-increment
in specific time unit experiences a linear growth, is proposed here.

   In our model, some spatial coordinates are given to the nodes to represent the research contents of papers (the differences of research contents are illustrated by the geometric distances between nodes).
   Besides, a simple spacetime, (2+1)-dimensional
 Minkowski spacetime of two spatial dimensions, along with one temporal dimension is considered in this model, so that the time characteristics of the nodes in citation networks can be modeled.
   The nodes in the model are uniformly and randomly sprinkled onto a cluster of concentric circles (the centers of which are on the time axis).
    In addition, the nodes on different circles are generated in different time units, while those in the same circle represent the papers published in the same issue. The number of nodes in a circle is a linearly increasing function of the circle's temporal coordinate.
   In the spacetime, nodes are identified by their locations $(R(t),\theta,t)$,  where $t$ is the generated time of the node, $R(t)$ is the radius of the circle born at time $t$, and $\theta$ is the angular coordinate.
 Considering that the radius $R(t)$ and the time $t$ are $1$-to-$1$ correspondence,  each node is identified by its location only with time coordinate $t$ and angular coordinate $\theta$. The edges in the model are linked according to the influence mechanism and the interdisciplinary citation mechanism, which are displayed in Fig~\ref{dia}. The influence zone of node $a_{1}$ contains node $a_{2}$, and thus a directed edge  is drawn  from node $a_{2}$ to node $a_{1}$ under a given probability. As node $a_{6}$ is    an ``interdisciplinary paper", it could cite node $a_{5}$, even though the influence zone of node $a_{5}$ does not contain $a_6$.

Supposing that a modeled network has $N(t)=mt$ papers ($m\in\mathbb{Z}^{+}$) published in the $t-$th unit of time $(t=1,2,...,T\in \mathbb{Z}^{+})$, including some interdisciplinary papers,   the generating process of the model is listed as follows.

\begin{enumerate}[Step 1]
  \item  Generate a new circle $C_{t}$ with radius $R(t)=N(t)/(2\pi\delta)$ ($\delta\in\mathbb{R}^{+}$) centered at point $(0,0,t)$ at each time $t=1,2,...T\in\mathbb{Z}^{+}$, sprinkle $N(t)$ nodes (papers) on it randomly and uniformly, and fix nodes with their coordinates, e.g. node $i$ with $(\theta_{i},t_{i})$.\\
  \item For each node with coordinate $(\theta,t)$, the influence zone (academic influence scope) of the node is defined as an interval of angular coordinate with center $\theta$
and arc-length   $D=\beta(\theta)/t^{\alpha}$, where  $\alpha\in(1,2)$   is used to tune the exponent of power-law tail of in-degree distribution, and $\beta(\theta)$  is used to make the in-degree distribution of papers published in each time unit have a power-law tail.   \\
  \item For node $i$ and node $j$, the coordinates of which are $(\theta_{i},t_{i})$ and $(\theta_{j},t_{j})$ respectively, if the distance of angular coordinates   $\Delta(\theta_{i},\theta_{j})=\pi-|\pi-|\theta_{i}-\theta_{j}||<|D_{i}|$ and $t_{j}>t_{i}$,  a directed edge is drawn from $j$ to $i$ under a probability $p$.\\
  \item Select $r (r\ll1)$ percent   nodes  as interdisciplinary papers to continually cite a number of existing papers randomly to make the reference lengths (out-degrees) of those papers to be random variables drawn from a power-law distribution $f(k)=k^{-\xi}$ $(k>k_{0})$.
\end{enumerate}

The function $\beta(\theta)$ in Step 2 is a  staircase function   of $\theta$
\begin{displaymath}
\beta(\theta)=\left\{\begin{array}{ll}
\beta_{0}, &\textrm{$\theta\in[0,\theta_{1}]$}\\
\beta_{0}+\lambda, &\textrm{$\theta\in[\theta_{1},\theta_{2}]$}\\
\vdots\\
\beta_{0}+(S-1)\lambda, &\textrm{$\theta\in[\theta_{S-1},2\pi]$},
\end{array}\right.
\end{displaymath}
where $\beta_{0}\in\mathbb{R}^{+}$, $\lambda>0$, $S\in \mathbb{Z}^+$,    and
 $[\theta_{0},\theta_{1}], ..., [\theta_{S-1},\theta_{S}]$ are a specific partition of $[0,2\pi]$ satisfying
$\Delta(\theta_{i+1},\theta_{i})=2\pi  { (\beta_{0}+i\lambda)^{-\eta}}/{\sum_{j=1}^ {S-1}(\beta_{0}+j\lambda)^{-\eta}}$, $i=0,2,...,S-1$,  $\eta>0$, $\theta_{0}=0$, $\theta_{S}=2\pi$, and the aging of the papers' influences is ignored here due to the short time span of the empirical data (around ten years) (Table~\ref{table1}).

 In this paper, the model is developed to fit Cit-HepTh and  Cit-HepPh~(Table~1).  The evolutionary trends of the monthly  numbers of papers in this two networks  are sufficiently fitted by  linear   functions~(Figs~\ref{fig1}a, \ref{fig1}b). To make the modeled time span (around ten years) and the modeled size of nodes match with the empirical data,  parameters are properly selected and listed in the end of Table~\ref{table1}. Especially, the unit of the   parameter $t$ is set as $2$ month~(Fig~\ref{fig1}c), while the rise rate of node-increment $m$ and the number of circles $T$ are set to be $15$ and $66$, respectively.

 \begin{figure}
 \begin{adjustwidth}{-2.25in}{0in}
\centering
  \includegraphics[height=3in,width=6in,angle=0]{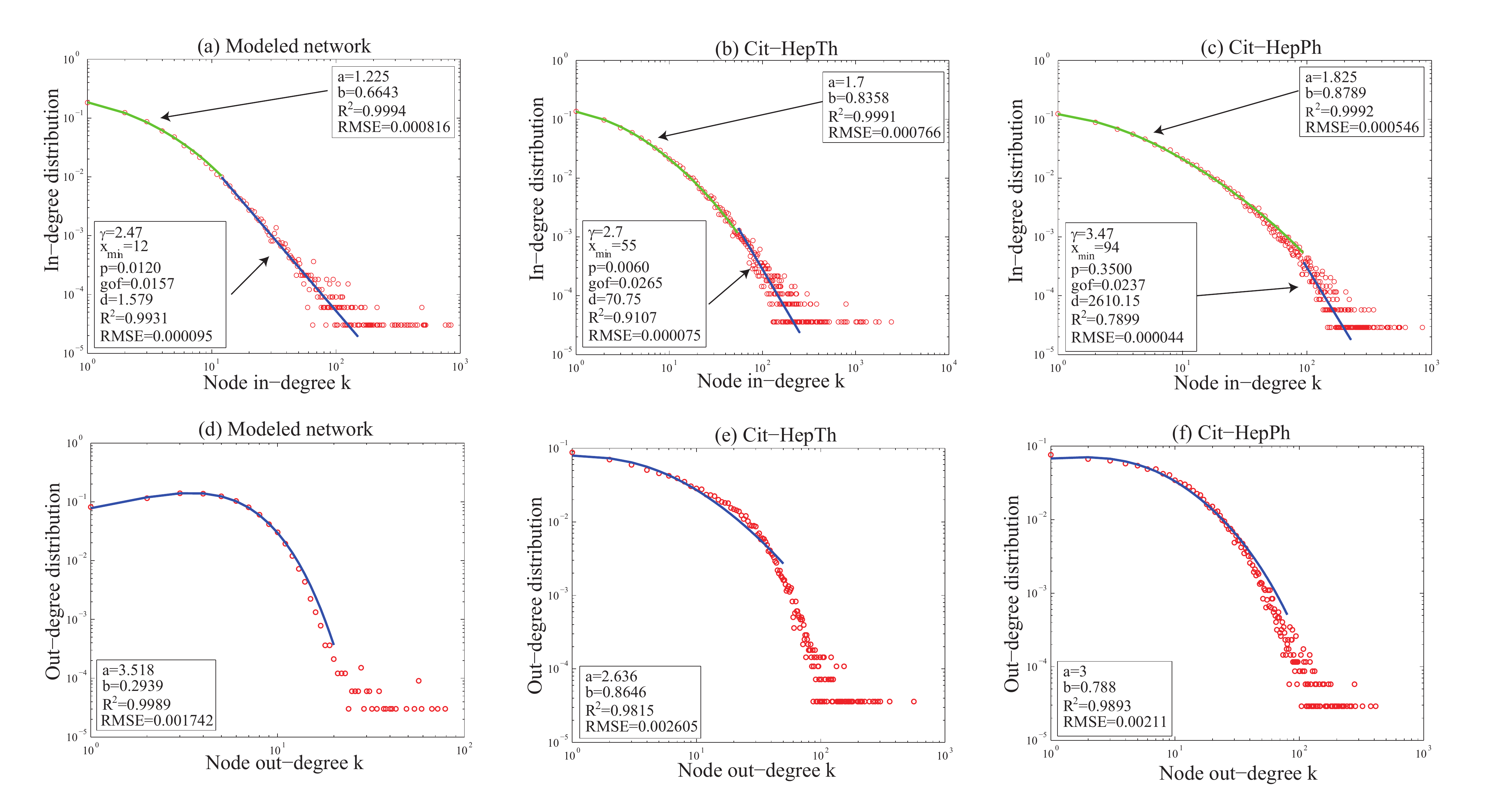}
\caption{\small\textbf{In- and out-degree distribution.} The fitting functions are the generalized Poisson distribution $f_{1}(k)=a(a+bk)^{k-1}e^{-a-bk}/k!$ for the foreparts of the in- and out-degree distributions, and the power-law distribution $f_{2}(k)=ck^{-\gamma}$ for the tails of the in-degree distributions in panels (a,b,c) (fitted by the method in Ref (\cite{liu18})). The root mean squared error (RMSE) and coefficient of determination $(R^{2})$ are used to measure the goodness of fits.}
\label{fig3}       
\end{adjustwidth}
\end{figure}

\section*{Out- and in- distribution}
The out-degree distributions of the empirical data (Table~\ref{table1}) take the form of  fat tails and curves in the forepart (Figs~\ref{fig3}e,\ref{fig3}f).  The curves in the forepart of the out-degree distributions can be well fitted by the generalized Poisson distribution. In reality, the behavior that paper $j$ cites paper $i$ is influenced by the number of the citations\cite{liu4,liu5,liu6,liu8} and the popularity of paper $i's$ author. At the same time, it can be viewed as a low-probability event (the reference length of paper $j$ is very small compared with the large number of papers). These settings are suitable for the use of the generalized Poisson distribution\cite{ref4}. Now the formulas of the forepart and tail of the out-degree distribution of the modeled network (Table~\ref{table1}) are derived to show how our model generates the similar curve and fat tail (Fig~\ref{fig3}d).

The edges in the model are linked according to the influence mechanism and the interdisciplinary citation mechanism.  Firstly,  the non-interdisciplinary paper $i$ with coordinate $(\theta_{i},t_{i})$ is considered. For prior node $j$, its coordinate is $(\theta_{j},t_{j})$, where $t_{j}<t_{i}$. If $\Delta(\theta_{i},\theta_{j})<\beta(\theta_{j})/t_{j}^{\alpha}$, node $i$ is located in the influence zone of node $j$. When $\beta(\theta_{j})/t_{j}^{\alpha}$ is small enough, $\beta(\theta_{i})\approx\beta(\theta_{j})$,  because $\beta(.)$ is a staircase function. Then the expected out-degree of node $i$ is as follows:
\begin{align}
k^{+}(\theta_{i},t_{i})=\sum_{t_{j}=1}^{t_{i}-1}\frac{\beta(\theta_{i})p}{t_{j}^{\alpha}}R(t_{j})\delta\approx\frac{m\beta(\theta_{i})pt_{i}^{2-\alpha}}{2\pi(2-\alpha)},
\end{align}
which is an increasing function of the temporal coordinate $t_{i}$. When $t_{i}$ is large enough, $\partial k^{+}(\theta_{i},t_{i})/\partial t_{i}\approx 0$, indicating that the reference length of the paper denoted by node $i$ is approximately a constant. This is in accordance with the actual situation that the reference length of papers cannot grow infinitely.

Since the process of sprinkling nodes follows  the Poisson point process, the actual out-degree of node $i$ is not exactly  equal to the expected out-degree. Therefore, in order to obtain the correct out-degree distribution, it is necessary to average the Poisson distribution\cite{liu19},
\begin{align}
p(k^{+}(\theta_{i},t_{i})=k)=\frac{1}{k!}(k^{+}(\theta_{i},t_{i}))^{k}e^{-k^{+}(\theta_{i},t_{i})},
\end{align}
which is the probability that node $i$ has out-degree $k$, with the temporal density $\rho(t_{i})$. In this model,
\begin{align}
\rho(t_{i})=\frac{mt_{i}}{\sum_{s=1}^{T}ms}\approx\frac{2t_{i}}{T^{2}}.
\end{align}
So the out-degree distribution is
\begin{align}
p_{\mathrm{non}}(k^{+}=k)&=\frac{1}{2\pi}\int_{0}^{2\pi}\int_{1}^{T}p(k^{+}(\theta_{i},t_{i})=k)\rho(t_{i})dt_{i}d\theta_{i}\nonumber\\
          &\approx\frac{1}{2\pi}\int_{0}^{2\pi}(\frac{m\beta(\theta_{i})p}{2\pi(2-\alpha)})^{k+\frac{\alpha}{2-\alpha}}\frac{e^{-\frac{m\beta(\theta_{i})p}{2\pi(2-\alpha)}}}{k!}d\theta_{i}.
\end{align}
It is a mixture poisson distribution similar to that of  the empirical data. Moreover, the curve in the forepart of the modeled in-degree distribution can be well fitted by the generalized Poisson distribution (Fig~\ref{fig3}d).

The interdisciplinary papers make the tail of the modeled out-degree distribution fat (Step 4) (Fig~\ref{fig3}d).
Thus, in combination with the non interdisciplinary papers, the out-degree distribution is
\begin{align}
p(k^{+}=k)=(1-r)p_{\mathrm{non}}(k^{+}=k)+rf(k),
\end{align}
where $r$ denotes the proportion of interdisciplinary papers, and $f(k)$ refers to the power-law distribution defined in Step 4.

The in-degree distributions of the empirical data have been investigated with the result showing that the curves in the forepart of the in-degree distributions can be well fitted by the generalized Poisson distribution (Figs~\ref{fig3}b,\ref{fig3}c). Actually, the citations of one paper are affected by the new papers of its authors, and the probability of one paper receiving citations (be selected from plenty of papers) is small  and not equal to that of other papers.  These are the conditions in which the generalized Poisson distribution can be applied\cite{ref4}. Besides, the in-degree distributions of the empirical data have a fat tail (Figs~\ref{fig3}b,\ref{fig3}c) which can be interpreted as a consequence of the cumulative advantage\cite{liu5,liu6,liu9,liu20,liu21}.
In this model, this phenomenon is caused by the highly cited papers with large influence zones.  Now, an expression of the forepart and tail of the modeled in-degree distribution is derived to show how our model generates the similar curve and fat tail(Fig~\ref{fig3}a).

For the modeled paper $i$, it can receive citations from the papers located inside or outside of its influence zone. Therefore, the expected in-degree of paper $i$ with coordinate $(\theta_{i},t_{i})$ is
\begin{align}\label{eq6}
k^{-}(\theta_{i},t_{i})&=\sum_{s=t+1}^{T}\frac{\beta(\theta_{i})p}{t^{\alpha}}R(s)\delta+r\sum_{s=t+1}^{T}\frac{2}{s-1}\nonumber\\
                       &\approx\frac{m\beta(\theta_{i})p}{4\pi t_{i}^{\alpha}}(T^{2}-t_{i}^{2})+2r\ln\frac{T-1}{t_{i}}.
\end{align}
When $t_{i}$ is small, the first item in formula~(\ref{eq6}) is larger than the  second item, so
\begin{align}\label{eq7}
k^{-}(\theta_{i},t_{i})\approx\frac{m\beta(\theta_{i})pT^{2}}{4\pi t_{i}^{\alpha}}.
\end{align}
Averaging the Poisson distribution, the in-degree distribution in the large in-degree region is
\begin{align}
p(k^{-}=k)&\approx\frac{1}{2\pi}\int_{0}^{2\pi}k^{-(1+\frac{2}{\alpha})}\int_{\frac{a_{2}}{T^{\alpha}}}^{a_{2}}\frac{e^{-\frac{(\tau-k+\frac{\alpha+2}{\alpha})^{2}}{2(k-\frac{\alpha+2}{\alpha})}}}{\sqrt{2\pi(k-\frac{\alpha+2}{\alpha})}}~d\tau~d\theta_{i},
\end{align}
where $a_{2}=m\beta(\theta_{i})pT^{2}/4\pi$, $\tau=a_{2}/t_{i}^{\alpha}$. The Laplace approximation  and the Stirling's approximation are used in this approximation. It can be proven that the integral term of $\tau$ is approximately independent of $k$. The derivation process is as follows:
\begin{align}
\frac{d}{dk}&\int_{\frac{a_{2}}{T}}^{a_{2}}\frac{e^{-\frac{(\tau-k+\frac{\alpha+2}{\alpha})^{2}}{2(k-\frac{\alpha+2}{\alpha})}}}{\sqrt{2\pi (k-\frac{\alpha+2}{\alpha})}}~d\tau
=\frac{e^{-\frac{(\tau-k+\frac{\alpha+2}{\alpha})^{2}}{2(k-\frac{\alpha+2}{\alpha})}}}{\sqrt{2\pi (k-\frac{\alpha+2}{\alpha})}}(1+\frac{\tau}{k-\frac{\alpha+2}{\alpha}})|_{\frac{a_{2}}{T}}^{a_{2}}\approx 0.
\end{align}
When the in-degree $k$ is large enough, which is satisfied by the small $t$ (formula~(\ref{eq7})), the integration is approximately equal to a constant. In this way, the modeled in-degree distribution in the large-$k$ has a power-law tail with exponent $1+2/\alpha$.

When $t_{i}$ is large,  the time derivative of the influence zone of paper $i$ is considered,
\begin{align}
(\frac{\beta(\theta_{i})}{t_{i}^{\alpha}})^{'}=-\alpha\beta(\theta_{i})t_{i}^{-\alpha-1}\approx 0.
\end{align}
It means that the influence zone of paper $i$ in this model is approximately a constant when $t_{i}$ is large. Hence, it is assumed that $\beta(\theta_{i})/t_{i}^{\alpha}\approx D$ (D is a constant). The expected in-degree of paper $i$ is
\begin{align}
k^{-}(\theta_{i},t_{i})\approx\frac{mDp}{4\pi}(T^{2}-t_{i}^{2})+2r\ln\frac{T-1}{t_{i}}.
\end{align}
And thus the in-degree distribution in the small in-degree region is
\begin{align}
p(k^{-}=k)&\approx\frac{4c\pi}{mDpT^{2}}\frac{(\frac{mDp}{4\pi}(T^{2}-1))^{k}e^{-\frac{mDp}{4\pi}(T^{2}-1)}}{k!}\nonumber\\
          &+\frac{(1-c)e^{-2\ln(T-1)}}{r}\frac{(2r\ln(T-1))^{k}e^{-2r\ln(T-1)}}{k!}.
\end{align}
It indicates that the in-degree distribution of the modeled network in the small in-degree region is a mixture Poisson distribution  similar to that of  the empirical data.  Also, the in-degree distribution of the modeled network in the small in-degree region can be well fitted by the generalized Poisson distribution (Fig~\ref{fig3}d).

\section*{Local clustering coefficient}
\begin{figure*}[!t]
\begin{adjustwidth}{-2.25in}{0in}
\centering
\includegraphics[height=1.5in,width=6in,angle=0]{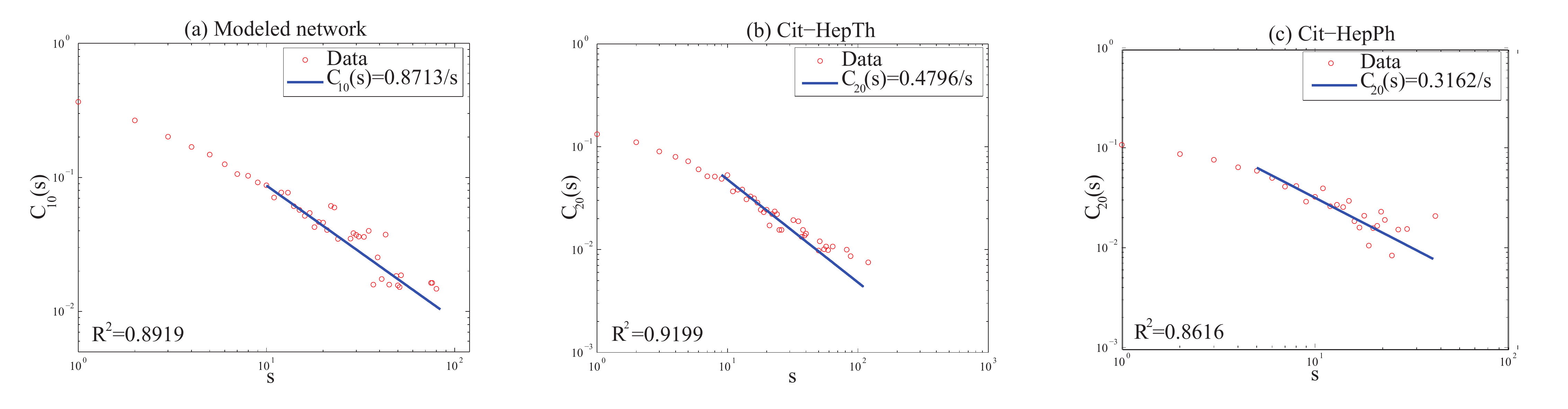}
\begin{center}
\caption{\small \textbf{The scaling relationships between the local clustering coefficients and the in-degrees of the citation networks in Table~\ref{table1}, compared with the theoretical prediction Eq~(\ref{eq14}).}  To reduce the noises caused by the random factors, the range of $k$ is divided into several intervals $[ls,l(s+1)]$, $l,s\in\mathbb{Z}^{+}$ to get $C_{l}(s)=\sum_{k=ls}^{l(s+1)}C(k^{-}=k)M(k)/\sum_{k=ls}^{l(s+1)}M(k)$, where $M(k)$ is the number of nodes with in-degree $k$.}
\label{fig4}
\end{center}
\end{adjustwidth}
\end{figure*}

The local clustering coefficient is equal to the probability that two vertices, both neighbors of the third vertex,  will be the neighbors of one another\cite{liu10,liu21}. Also, it is found that the highly cited papers often have low local clustering coefficients in empirical data (Figs~\ref{fig4}b,\ref{fig4}c). Due to the short time span of the empirical data (only ten years) (Table~\ref{table1}), the highly cited papers get  lots of citations from the new published papers with few citations.  In this paper, the highly cited papers are considered and the formula of the relation between the highly cited papers and their local clustering coefficients is derived to show how well our model fits the tail of the local clustering coefficient.

  Suppose $i$  is a highly cited paper and $t_{i}$ is small enough. Paper $j$ and paper $l$ are the new published papers, which are the neighbors of paper $i$. If $j$ has coordinate $(\theta_{j},t_{j})$, a reasonable assumption is made that the overlap of the influence zones of $i$ and $j$ in circle $C_{t_{c}}$ ($t_{c}$ is the current time) is approximately $\beta(\theta_{j})/t_{j}^{\alpha}$ because of the small $t_{i}$ and large $t_{j}$. Particularly, if the connection probability $p$ equals to 1, the probability that paper $l$ is the common neighbor of paper $i$ and paper $j$ is approximately equal to $\beta(\theta_{j})t_{i}^{\alpha}/\beta(\theta_{i})t_{j}^{\alpha}$. Thus, for the general connection probability $p$, the conditional probability $p(l\mapsto j|l\mapsto i, j\mapsto i)=p\beta(\theta_{j})t_{i}^{\alpha}/\beta(\theta_{i})t_{j}^{\alpha}$. The effect of the interdisciplinary papers is ignored here owing to the low probability that paper $l$ is an interdisciplinary paper and connects to paper $i$ and paper $j$ simultaneously.  Summing over the possible values of $t_{j}$, it can be found that
\begin{align}\label{eq13}
C(\theta_{i},t_{i})=\frac{\int_{t_{i}}^{T}p^{2}\frac{\beta(\theta_{j})t_{i}^{\alpha}}{\beta(\theta_{i})t_{j}^{\alpha}}\sigma~dt_{j}}{\int_{t_{i}}^{T}p\sigma~dt_{j}}=\frac{2p\beta(\theta_{j})t_{i}^{\alpha}}{(2-\alpha)\beta(\theta_{i})}\frac{T^{2-\alpha}-t_{i}^{2-\alpha}}{T^{2}-t_{i}^{2}},
\end{align}
where $\sigma=m\beta(\theta_{i})t_{j}/2\pi t_{i}^{\alpha}$ denotes the number of the papers in the influence zone of paper $i$ at time $t_{j}$.

Since paper $i$ is a highly cited paper, the papers citing $i$ dominate the neighbors of $i$ and the effect of papers cited by $i$ can be ignored. Moreover, the expected in-degree of the highly cited paper $i$ is  $m\beta(\theta_{i})pT^{2}/4\pi t_{i}^{\alpha}$. By substituting it into Eq~(\ref{eq13}), we get
\begin{align}\label{eq14}
C(k^{-}(\theta_{i},t_{i})=k)\approx\frac{m\beta(\theta_{j}) p^{2}T^{2-\alpha}}{2\pi(2-\alpha)k}\propto\frac{1}{k},
\end{align}
which is inversely proportional to the in-degree $k$ of paper $i$. Thus, the local clustering coefficient of the highly cited paper $i$ in this model is also small. To show the similarity more clearly, the range of $k$ is divided into equal small intervals and $C(k^{-}(\theta_{i},t_{i})=k)$ for each interval is also averaged to reduce the noises caused by random factors (Fig~\ref{fig4}).

\section*{In- and out-Assortativity}
It can  be seen  that the highly cited papers tend to cite the same highly cited papers in the empirical data, which means that they are in-assortative (Table~\ref{table1}). Moreover, it is intuitive that researchers are often wild about tracing back to hot topics. If the topics of papers have great research value, numerous researchers will focus on them and publish a large number of  papers that will cite each other. As a result, these papers become highly cited as well. The empirical data are also out-assortative (Table~\ref{table1}), which refers to the tendency of papers to cite other papers with similar out-degrees to themselves. Actually, the researchers  often put emphasis on  the new published papers that have novel contents.

\begin{figure*}[!t]
\begin{adjustwidth}{-2.25in}{0in}
\centering
\includegraphics[height=1.5in,width=6in,angle=0]{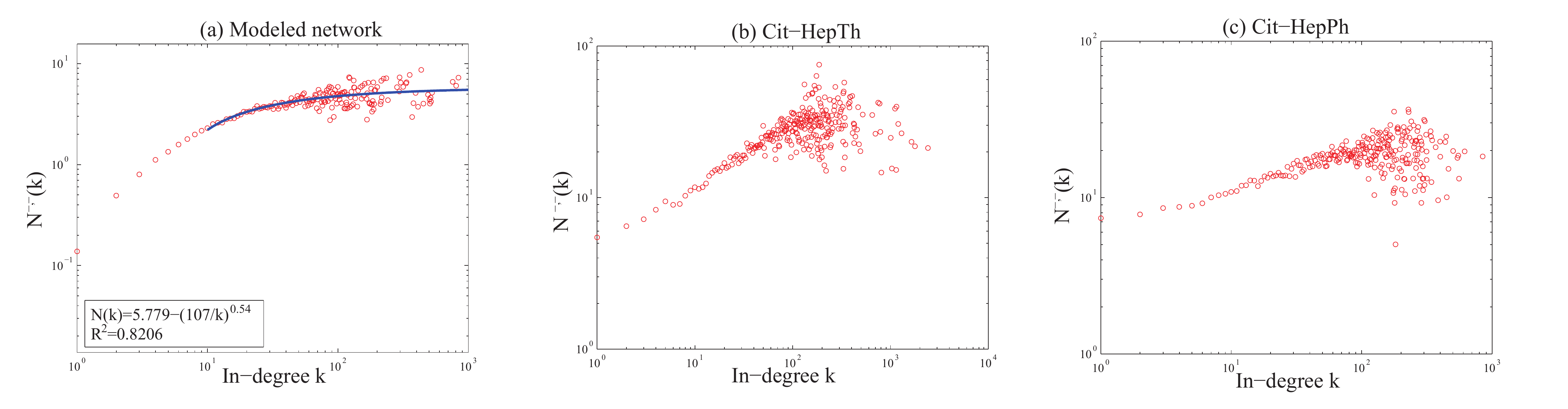}
\begin{center}
\caption{\small \textbf{The scaling relation $N^{-,-}(k)$ between the in-degree $k$ and the mean in-degree of the neighbors pointing to the nodes with in-degree $k$.} Panel (a) shows $N^{-,-}(k)$ (fitted by Eq~(\ref{eq16})) for the modeled network in Table~\ref{table1}. Panels $(b,c)$ show $N^{-,-}(k)$ for two empirical data in Table~\ref{table1}.
}
\label{fig5}
\end{center}
\end{adjustwidth}
\end{figure*}

\begin{figure*}[!t]
\begin{adjustwidth}{-2.25in}{0in}
\centering
\includegraphics[height=1.5in,width=6in,angle=0]{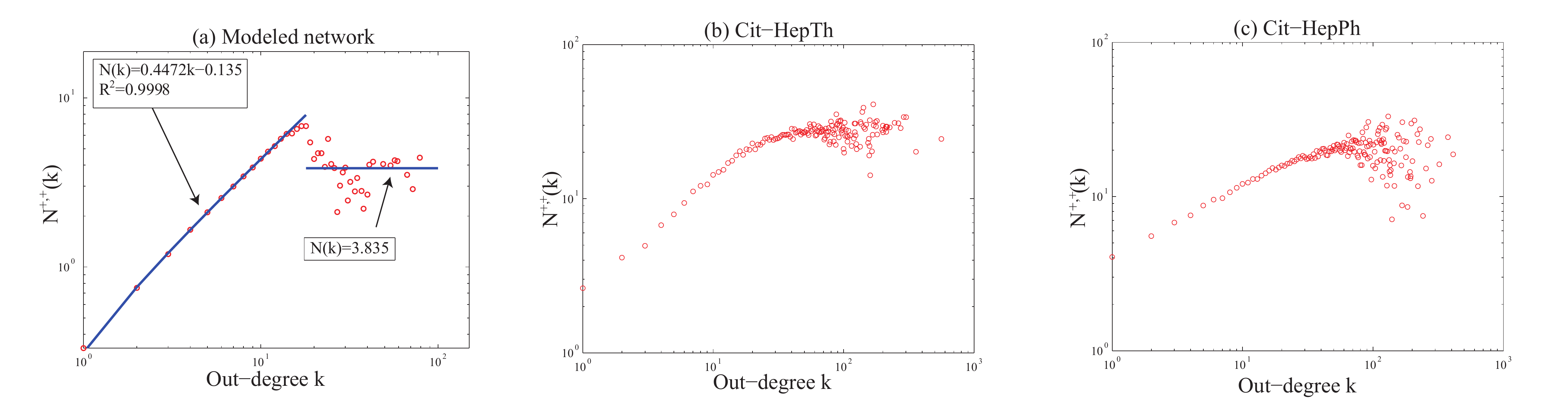}
\begin{center}
\caption{\small \textbf{The scaling relation $N^{+,+}(k)$ between the out-degree $k$ and the mean out-degree of the neighbors pointed at by the nodes with out-degree $k$.} Panel(a) shows $N^{+,+}(k)$  (fitted by Eq~(\ref{eq17}) and Eq~(\ref{eq19})) for the modeled network in Table~\ref{table1}.  Panels $(b,c)$ show $N^{+,+}(k)$ for two empirical data in Table~\ref{table1}.
}
\label{fig6}
\end{center}
\end{adjustwidth}
\end{figure*}

Our model also has these two properties. To show the performance of this model, the formulas of the scaling relations between the in- and out-degree of a node and the mean in- and out-degree of the neighbors pointing to and pointed at by the node are derived. The relations are denoted by $N^{-,-}$ and $N^{+,+}$, respectively.

For node $i$, the coordinate of which is $(\theta_{i},t_{i})$, the in-degrees of all nodes pointing to it are averaged, and found
\begin{align}\label{eq15}
N^{-,-}(\theta_{i},t_{i})&=\frac{\int_{t_{i}}^{T}(\frac{m\beta(\theta_{i}) p}{4\pi s^{\alpha}}(T^{2}-s^{2})+2\xi\ln\frac{T-1}{s})\frac{m\beta(\theta_{i}) ps}{2\pi t_{i}^{\alpha}}~ds}{\int_{t_{i}}^{T}\frac{m\beta(\theta_{i}) ps}{2\pi t_{i}^{\alpha}}~ds}\nonumber\\
          &\approx\frac{m\beta(\theta_{i}) p}{2\pi(2-\alpha)}(T^{2-\alpha}-t_{i}^{2-\alpha}).
\end{align}
In this formula, some approximations are made, and they hold for small $t_{i}$, meaning that formula~(\ref{eq15}) can only fit the tail of the scaling relation.
 Therefore, substituting the in-degree $k$  of node $i$ born early into formula~(\ref{eq15}), we get 
\begin{align}\label{eq16}
N^{-,-}(k^{-}(\theta_{i},t_{i})=k)\approx u-(\frac{v}{k})^{\frac{2-\alpha}{\alpha}},
\end{align}
where $u$ and $v$ are constants. Formula~(\ref{eq16}) is an increasing function, suggesting that the model is in-assortative. In addition, when in-degree $k$ is large enough,  $N^{-,-}(\theta_{i},t_{i})$ is approximately equal to a constant (Fig~\ref{fig5}a). It is close to the actual situation that a hot topic will be out of fashion.

Then $N^{+,+}(\theta_{i},t_{i})$ is considered.
If node $i$ is not an interdisciplinary paper, we could get
\begin{align}\label{eq17}
N^{+,+}(\theta_{i},t_{i})&\approx\frac{\int_{1}^{t_{i}}\frac{m\beta(\theta_{i})ps^{2-\alpha}}{2\pi(2-\alpha)}\frac{m\beta(\theta_{i})ps^{1-\alpha}}{2\pi}~ds}{\int_{1}^{t_{i}}\frac{m\beta(\theta_{i})ps^{1-\alpha}}{2\pi}~ds}
          =\frac{m\beta(\theta_{i})p}{2\pi(4-2\alpha)}(t_{i}^{2-\alpha}+1).
\end{align}
Substituting the expected out-degree of node $i$ into formula~(\ref{eq17}), we get
\begin{align}\label{eq18}
N^{+,+}(k^{+}(\theta_{i},t_{i})=k)=\frac{1}{2}k+\frac{m\beta(\theta_{i})p}{2\pi(4-2\alpha)}\propto\frac{1}{2}k+c,
\end{align}
where $c$ is a constant.  Formula~(\ref{eq18}) is an increasing function about out-degree $k$,  whereas $N^{+,+}(\theta_{i},t_{i})$ will not satisfy the result given by Formula~(\ref{eq18}) (Fig~\ref{fig6}a) if the out-degree of node $i$ is large, which shows that most nodes with large out-degrees represent interdisciplinary papers (Fig~\ref{fig3}d) . If node $i$ is an interdisciplinary paper, the nodes that are pointed at by it are also interdisciplinary papers due to the out-assortativity of the model. So
\begin{align}\label{eq19}
N^{+,+}(k^{+}(\theta_{i},t_{i})=k)\approx\frac{1-\gamma}{2-\gamma}k_{0}\propto C,
\end{align}
where $C$ is a constant. It indicates that the average out-degree of nodes pointed at by node $i$ fluctuates around a constant (Fig~\ref{fig6}a). Meanwhile, the model is out-assortative, as the number of interdisciplinary papers in the model is small.

\section*{Conclusion}
   A model of scientific citation networks with linearly growing node-increment is proposed, in which the influence mechanism and the interdisciplinary citation mechanism are involved. Under appropriate parameters, the formula of the modeled network's  in-degree distribution is derived, and it shows a similar behavior to the empirical data in the small in-degree region and a power-law tail in the large in-degree region.  Different from most previous models that just study the forepart of the out-degree distribution of the empirical data, this model also captures the fat tails. The  model can also predict some other typical statistical features like clustering,  in- and out-assortativity, giant component and clear community structure. For example, it vividly characterizes the academic influence power of papers by geometric zones, and interprets the power-law tails of  citation networks' in-degree distributions  by the papers' inhomogeneous influence power. Therefore, it is believed that this model is a suitable geometric tool to study the citation networks. However, some shortcomings still need to be overcome in future work: how to design a mechanism to characterize the citations of the interdisciplinary papers rather than randomly and uniformly select the existing papers; and how to model the out-degree distribution better.


\linenumbers
\section*{Acknowledgments}
The authors would like to thank Pengyuan Zhang, Zonglin Xie and Han Zhang for helpful
discussions and Dan Zhuge for proofreading this paper.

\section*{Author Contributions}
Conceived and designed the experiments: QL. Performed the experiments: QL.
Analyzed the data: QL ZX ED. Contributed reagents/materials/analysis tools: QL ZX ED.
Wrote the paper: QL ZX ED JL.

\nolinenumbers

%
%
%

\end{document}